\documentclass[11pt,journal,onecolumn]{IEEEtran}

\usepackage{multicol}
        
\usepackage{amssymb,amsmath,amscd,amsfonts,amstext}
\usepackage{graphicx}
\usepackage{epsfig,color}
\usepackage{colortbl}
\usepackage{hyperref}

\usepackage{graphicx}
\usepackage{amsmath}
\usepackage{amssymb}
\usepackage{epsfig,color}

\newtheorem{lemma}{Lemma}

\newcommand{\A}{{\bf A}}
\newcommand{\C}{{\bf C}}                                                                                                                            

\newcommand{\I}{{\bf I}}

\newcommand{\E}{{\bf E}}

\newcommand{\G}{{\bf G}}

\newcommand{\x}{{\bf x}}     

\newcommand{\s}{{\bf s}}

\renewcommand{\baselinestretch}{2}          

\begin{document}
\title{Exact Conditional and Unconditional Cram\`er-Rao Bounds for Near Field Localization}
\author{Youcef Begriche$^1$\thanks{$^1$ TELECOM ParisTech, TSI Department, France. e-mail:Youcef.Begriche@ieee.org}, Messaoud Thameri$^{1}$\thanks{$^1$ TELECOM ParisTech, TSI Department, France. e-mail:thameri@telecom-paristech.fr} and  Karim Abed-Meraim$^{2}$\thanks{$^2$ PRISME Laboratory, Polytech'Orl\'eans, Orl\'eans University, France. e-mail: karim.abed-meraim@univ-orleans.fr }}

\maketitle

\begin{abstract}
This paper considers the Cramer-Rao lower Bound (CRB) for the source localization problem in the near field.
More specifically, we use the exact expression of the delay parameter for the CRB derivation and show how this `exact CRB' can be significantly different from the one given in the literature and based on an approximate time delay expression (usually considered in the Fresnel region).
This CRB derivation  is then generalized by considering the exact expression of the received power profile (i.e., variable gain case) which, to our best knowledge, has been ignored in the literature.
Finally, we exploit the CRB expression to introduce the new concept of Near Field Localization (NFL) region  for a target localization performance associated to the application at hand. We illustrate the usefulness of the proposed CRB derivation and its developments as well as the NFL region concept through numerical simulations in different scenarios.
\end{abstract}

\newpage
\section{Introduction}
\label{Introduction}
Sources localization problem has been extensively studied in the literature but most of the research works are dedicated to the far field case, e.g., \cite{H_Krim_M_Viberg, livre_far_field}. 

In this paper, we focus on the situation where the sources are in a near field region which occurs when the source ranges to the array are not `sufficiently large' compared with the aperture of the array system \cite{ref2,ref9}. Indeed, this particular case has several practical applications including speaker localization and robot navigation \cite{refG4,refG6}, underwater source localization \cite{refG5}, near field antenna measurements \cite{NFM1}, \cite{NFM2} and certain biomedical applications, e.g., \cite{refG7}. Recently, some works considered both far filed and near filed localization \cite{loc_NFFF1}, \cite{loc_NFFF2} where the authors propose different methods to achieve a better localization performance when the source moves from far field to near field and vice versa.

More specifically, this paper is dedicated to the derivation of the Cramer Rao Bound expressions for different signal models and their use for the better understanding of this particular localization problem. CRB derivation for the near field case has already been considered in the literature \cite{ref2,ref4,kopp,elkorso}. In \cite{ref2,ref4}, the exact expression of the time delay has been used to derive the unconditional CRB in matrix form, i.e., expressed and computed numerically as the inverse of the Fisher Information Matrix (FIM). In \cite{kopp}, the conditional CRB based on an approximate model (i.e., approximate time delay expression as shown in Section \ref{data_model}) is provided.

Recently, EL Korso et al. derived analytical expressions of the conditional and unconditional CRB for near field localization based on an approximate model \cite{elkorso}. In the latter work both conditional and unconditional CRB of the angle parameter are found independent from the range value and are equal to those of far field region.

In our work, we propose first, to use the exact time delay expression for the derivation of the conditional and unconditional CRB and provide closed form formula that are compared to those given in \cite{elkorso}. The development (i.e., Taylor expansion) of the exact CRB allows us to highlight many interesting features including: (i) a more accurate approximate CRB for the exact model as compared to the  CRB based on an approximate model, (ii) and a detailed analysis of the source range parameter effect on its angle estimation performance\footnote{Part of this work has been published in \cite{CRB_ISSPA2012}}.

Secondly, we take into consideration the spherical form of the wavefront into the power profile. Indeed, in the near field case, the received power is variable from sensor to sensor which should be taken into account in the data model. By considering such variable gain model, we generalize the previous CRB analysis and investigate the impact of the gain variation onto the localization performance limit.

Finally, we propose to exploit the exact CRB expression to specify the 'near field localization region' based on a desired localization performance. In that case, the 'near field localization region' is shown to depend not only on the source range parameter and array aperture but also on the sources SNR and observation sample size.

The paper is organized as follows: Section \ref{problem} introduces the data model and formulates the main paper objectives. Section \ref{equal_gain}, the exact conditional and unconditional CRB derivations and their Taylor expansions are provided for the equal gain case.
Section \ref{variable_gain} generalizes the previous analysis to the variable gain model and highlights the differences between equal and variable gain cases.
In Section \ref{near_field}, we introduce the concept of near field localization region and illustrate its usefulness through specific examples.
Section \ref{Simulation} is dedicated to simulation experiments while Section \ref{conclusion} is for the concluding remarks.
\section{Problem formulation}
\label{problem}

\subsection{Data model}
\label{data_model}
\begin{figure}%
\centerline{\includegraphics[width=5cm]  {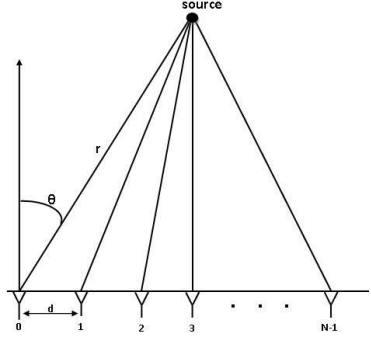}}%
\caption{Near field source model}%
\label{fig:model}%
\end{figure}

In this paper, we consider a uniform linear array with $N$ sensors receiving a signal, emitted from one source located in the near field region, and corrupted by circular white Gaussian noise $v_n$ of covariance matrix $\sigma^2\I_N$. The $n^{th}$ array output, $n=0,\cdots,N-1$, is expressed as 
\begin{equation}
	x_{n}(t) = s(t)e^{j\tau_{n}} + v_{n}(t) ~~~~~~ t=1,\cdots,T
\label{eq:outputs_eg}
\end{equation}
where $s(t)$ is the emitted signal. The exact expression of the time delay\footnote{The first sensor, $n=0$, is considered for the time reference.}  $\tau_n$ is given by
\begin{equation}
	\tau_{n}=\frac{2\pi r}{\lambda}\left(\sqrt{1+\frac{n^{2}d^{2}}{r^{2}}-\frac{2nd \sin\theta}{r}}-1\right)
\label{eq:exact_tau}
\end{equation}
where $d$ is the inter-element spacing, $\lambda$ is the propagation wavelength and ($r, \theta$) are the polar coordinates of the source as shown in Fig. \ref{fig:model}.
In the literature, the near field region (also called Fresnel region) is given by \cite{fresnel}
\begin{equation}
	0.62\left(\frac{d^3(N-1)^3}{\lambda}\right)^{\frac{1}{2}}<r<2\frac{d^2(N-1)^2}{\lambda}
\label{eq:Fresnel_region}
\end{equation}

Most existing works on near field source localization consider the following approximation of the time delay expression to derive simple localization algorithms as well as CRB expressions, e.g., \cite{Grosicki} \cite{elkorso}
\begin{equation}
	\tau_n = -2\pi \frac{dn}{\lambda}\sin(\theta) + \pi \frac{d^2n^2}{\lambda} \frac{1}{r}\cos^2(\theta) + o\left(\frac{d}{r}\right)
\label{eq:approximated_tau}
\end{equation}

In the sequel, the source signal will be treated either as deterministic (conditional model) or stochastic (unconditional model). Indeed, in the array processing, both models can be found:
\begin{enumerate}
	\item Conditional model in which the source signal is assumed deterministic but its parameters are unknown.
	\item Unconditional model in which we assume that the source signal is random. In our case, we will assume $\s(t)$ to be a complex circular Gaussian process with zero mean and unknown variance $\sigma_s^2$.
\end{enumerate}
\subsection{Objectives}
\label{objectives}
\begin{enumerate}
	\item  Based on the exact expression of the time delay $\tau_n$ in equation (\ref{eq:exact_tau}), we aim to derive the exact conditional and unconditional CRB and compare them with the existing CRB given in \cite{elkorso} and derived from the approximate model in (4).
	\item  The distance from the source to the $n^{th}$ sensor $d_n$ is a function of the sensor position (i.e., sensor index) according to
	\begin{eqnarray}
  d_n & = & \sqrt{r^2-2ndr\sin(\theta)+n^2d^2}  \nonumber \\
      & = & r\sqrt{1-2\frac{nd}{r}\sin(\theta)+\left(\frac{nd}{r}\right)^2}
  \label{eq:source_sensor}
  \end{eqnarray}
	where $r$ is the distance from the source to the reference sensor (i.e., for $n=0$, $d_0 = r$).	
	Hence, the received power profile is variable from sensor to sensor. In the far field case, this variation is negligible but not necessarily in the near field context. To the best of our knowledge, this point has not been taken into consideration in the existing literature. We would like to investigate the impact of such gain variation into the localization performance limit.
	\item  The near field region has been so far assimilated to the Fresnel region which depends on the antenna size and the signal wavelength only.  However, the localization performance depends on other system parameters (SNR, sample size, angle position, $\cdots$) for which reason we introduce the concept of near field localization region where the localization error is upper bounded by a desired threshold value depending on the considered application.
  \end{enumerate}
  
\section{Conditional and unconditional CRB derivation with Equal Gain (EG)}
\label{equal_gain}
Here, we do not consider the variation of the power profile which is equivalent to assuming that all sensors have equal gain as shown in equation (1). The goal of this Section is to derive the conditional and unconditional CRB for the angle and range parameter estimation of near field source and to make comparison with the results presented in \cite{elkorso}. 

\subsection{Exact conditional CRB  with EG}
\label{ExactCRBCeg}
In this Section, we consider the source signal as deterministic according to the model

$$ s(t) = \alpha(t) e^{j(2\pi f_{0}t + \psi(t))}$$
where $f_{0}$ is the known carrier frequency while $\alpha(t)$ and $\psi(t)$ are the unknown amplitude and phase parameters of the source signal. Under the data model assumption of Section \ref{data_model}, we derive next the exact deterministic (i.e., conditional) CRB for the location source parameter estimation.

\subsubsection{Exact conditional CRB derivation}
In this deterministic case, the log-likelihood function of the observations is given by
$$l(\boldsymbol{\xi})=-NT\ln\pi - NT\ln \sigma^{2} - \frac{1}{\sigma^{2}}\left\|\boldsymbol{x}-\boldsymbol{\mu}\right\|^{2}$$
where 
\begin{eqnarray*}
\boldsymbol{x}      &=& [\x^T(1)\cdots ,\cdots,\x^T(T)]^{T}\\ 
\x^T(t)             &=& [x_0(t),\cdots,x_{N-1}(t)]^T \\
\boldsymbol{\mu}    &=& [s(1) \boldsymbol{a}^{T}(\theta,r),\cdots,s(T)\boldsymbol{a}^{T}(\theta,r)]^{T}\\
\boldsymbol{a}(\theta,r) &=& [1, e^{j  \tau_{1}(\theta,r)},\cdots, e^{j  \tau_{N-1}(\theta,r)}]^{T}\\
\boldsymbol{\xi}    &=& [\theta,r,\boldsymbol{\Psi}^{T},\boldsymbol{\alpha}^{T},\sigma^{2}]^{T}\\
\boldsymbol{\Psi}   &=& [\psi(1),\cdots, \psi(T)]^{T}\\
\boldsymbol{\alpha} &=& [\alpha(1), \cdots, \alpha(T)]^{T}
\end{eqnarray*}
where $\left\|.\right\|$ refers to the Frobenius norm and $[.]^{T}$ is the transpose operator.

The CRB is equal to the inverse of the Fisher Information Matrix (FIM) defined by
\begin{equation}
	[\mbox{FIM}(\boldsymbol{\xi})]_{i,j}=E\left(\frac{\partial l(\boldsymbol{\xi})}{\partial \xi_{i}}\frac{\partial l(\boldsymbol{\xi})}{\partial \xi_{j}}\right)
\end{equation}
the latter is given in our particular case by
\begin{equation}
	[\mbox{FIM}(\boldsymbol{\xi})]_{i,j}=\frac{NL}{\sigma^{4}} \frac{\partial{\sigma^{2}}}{\partial{\xi}_{i}} \frac{\partial{\sigma^{2}}}{\partial\xi_{j}}+ \frac{2}{\sigma^{2}} \mathcal{R}e \{\frac{\partial{\boldsymbol{\mu}^{H}}}{\partial{\xi}_{i}} \frac{\partial{\boldsymbol{\mu}}}{\partial{\xi}_{j}}\}
\label{eq:FIM_particular}
\end{equation}
where $\mathcal{R}e\{.\}$ refers to the real part of a complex valued entity.
The FIM is given by the block diagonal matrix as
\begin{displaymath}
 {\mbox{FIM}} = {\left[ \begin{array}{c  c   c}
                         {\mathcal{\textbf {\textit{Q}}}}&   &{\bf 0} \\
                                      & \frac{2N}{\sigma^2}{\bf I}_T &  \\
                                  {\bf 0}&  & \frac{NT}{\sigma^4}
                          \end{array} \right]}
\end{displaymath}

Indeed, it is shown in \cite{elkorso}, that the FIM of the desired location parameters is decoupled from the noise variance $\sigma^{2}$ and the sources magnitude 
parameters $\boldsymbol{\alpha}$. Consequently, the CRB of the range and angle parameters is equal to the $2\times 2$ top left sub-matrix of the inverse matrix
\begin{displaymath}
	{\mathcal{\textbf {\textit{Q}}}}^{-1} = {\left[ \begin{array}{c  c |  c}
																									f_{\theta \theta} & f_{\theta r}  & \textbf {f}_{\theta \psi} \\
																									f_{r \theta} & f_{r r} & \textbf {f}_{r \psi} \\
																									\hline
																									\textbf{f}_{\psi \theta} & \textbf{f}_{\psi r} & \textbf{F}_{\psi \psi}
																										\end{array} \right]}^{-1}
\end{displaymath}
where 
\begin{eqnarray*}
f_{\theta \theta} &=& 2TD_{\mbox{SNR}}\left\|\dot{\boldsymbol{\tau}}_{\theta}\right\|^{2} 															\\
f_{r r}           &=& 2TD_{\mbox{SNR}}\left\|\dot{\boldsymbol{\tau}}_{r}\right\|^{2}      															\\
f_{r \theta}      &=& f_{\theta r}=2TD_{\mbox{SNR}} (\dot{\boldsymbol{\tau}}^{H}_{\theta}\dot{\boldsymbol{\tau}}_{r})  \\
D_{\mbox{SNR}}           &=& \frac{{\left\|\boldsymbol{\alpha}\right\|}^{2}}{T\sigma^{2}} \ \ \ \text{(deterministic \mbox{SNR})}   \\
\dot{\boldsymbol{\tau}}_{\theta}&=& \left[\frac{\partial{\tau_0}}{\partial{\theta}},\ldots,\frac{\partial{\tau_{N-1}}}{\partial{\theta}}\right]^{T} \\
\dot{\boldsymbol{\tau}}_{r}&=& \left[\frac{\partial{\tau_0}}{\partial{r}},\ldots,\frac{\partial{\tau_{N-1}}}{\partial{r}}\right]^{T}
\end{eqnarray*}

Moreover, the vectors of size $T\times1$, $\textbf {f}_{\psi \theta}$, $\textbf {f}_{\theta \psi}^{T}$, $\textbf {f}_{\psi r}$ and $\textbf {f}_{r \psi}^{T}$ are given by
\begin{eqnarray*}
\textbf {f}_{\psi \theta} &=& \textbf {f}_{\theta \psi}^{T}=\frac{2}{\sigma^{2}} (\boldsymbol{1}^{T}_{N} \dot{\boldsymbol{\tau}}_{\theta}) (\boldsymbol{\alpha} \odot  \boldsymbol{\alpha}) \\
\textbf {f}_{\psi r}      & =& \textbf {f}_{r \psi}^{T} =\frac{2}{\sigma^{2}} (\boldsymbol{1}^{T}_{N} \dot{\boldsymbol{\tau}}_{r})(\boldsymbol{\alpha} \odot  \boldsymbol{\alpha})
\end{eqnarray*}
where $\odot$ represents the Hadamard product and ${\bf 1}_N$ is the all-one vector of size $N\times 1$. Finally, matrix \textbf {F}$_{\psi \psi}$ of size $T \times T$, is given by
\begin{equation*}
\textbf {F}_{\psi \psi}=\frac{2N}{\sigma^{2}}diag(\boldsymbol{\alpha} \odot  \boldsymbol{\alpha})
\end{equation*}
which is the diagonal matrix formed by vector $\boldsymbol{\alpha} \odot  \boldsymbol{\alpha}$.

By using the block structure of the FIM and the Schur's matrix inversion lemma \cite{schur}, we obtain the closed form expression of the CRB given by the following lemma:
\begin{lemma}
The non-matrix expressions of the exact conditional CRB in the equal gain case for a source in the near field, for $N \geq 3$ and $\theta \neq \pm\frac{\pi}{2}$, are given by
\begin{center}
\begin{tabular}{|c c|}
\hline
$\mbox{CRB}^{c}_{eg}(r)    =\left(\frac{1}{2T D_{\mbox{SNR}}}\right)\frac{E_{eg}(\theta)}{E_{eg}(\theta)E_{eg}(r)-E_{eg}(r,\theta)^{2}}$ &  \stepcounter{equation}\thetag{\theequation}\label{eq:exctCRBeg_r}\\%
$\mbox{CRB}^{c}_{eg}(\theta)   =\left(\frac{1}{2T D_{\mbox{SNR}}}\right)\frac{E_{eg}(r)}{E_{eg}(\theta)E_{eg}(r)-E_{eg}(r,\theta)^{2}}$&  \stepcounter{equation}\thetag{\theequation}\label{eq:exctCRBeg_th}\\
$\mbox{CRB}^{c}_{eg}(r,\theta) =\left(\frac{1}{2T D_{\mbox{SNR}}}\right)\frac{E_{eg}(r,\theta)}{E_{eg}(\theta)E_{eg}(r)-E_{eg}(r,\theta)^{2}} $&  \stepcounter{equation}\thetag{\theequation}\label{eq:exctCRBeg_rth}\\
\hline
\end{tabular}
\end{center}
where $\mbox{CRB}^{c}_{eg}(r,\theta)$ is the non diagonal entry of the considered $2\times 2$ CRB matrix (it represents the coupling between the $2$ parameters) and 
\begin{eqnarray*}
E_{eg}(\theta) &=&\left\|\dot{\boldsymbol{\tau}}_{\theta}\right\|^{2} - \frac{1}{N}({\bf 1}_N^{T} \dot{\boldsymbol{\tau}}_{\theta})^{2}               \\
E_{eg}(r) &=&\left\|\dot{\boldsymbol{\tau}}_{r}\right\|^{2} - \frac{1}{N}({\bf 1}_N^{T}\dot{\boldsymbol{\tau}}_{r})^{2}                              \\
E_{eg}(r,\theta) &=&\dot{\boldsymbol{\tau}}^{H}_{\theta}\dot{\boldsymbol{\tau}}_{r} - \frac{1}{N}({\bf 1}_N^{T} \dot{\boldsymbol{\tau}}_{\theta})({\bf 1}_N^{T} \dot{\boldsymbol{\tau}}_{r})  
\end{eqnarray*}
\label{lm:exact_CRBCeg}
\end{lemma}
\textbf{Proof}: The proof of this lemma can be deduced directly from lemma \ref{lm:exact_CRBCvg} proof given in appendix \ref{App1}.
\subsubsection{Taylor Expansion (TE) of the CRB}
\label{taylor}
This section aims to highlight the effects of some system parameters on the localization performance  and to better compare our exact CRB with the CRB given in \cite{elkorso}. For that, we propose to use a Taylor expansion of the expressions of lemma \ref{lm:exact_CRBCeg}. For simplicity, we omit the details of the cumbersome (but straightforward) derivations and present only the final results in the following lemma:
\begin{lemma}
The Taylor expansions of the exact CRB expressions of lemma \ref{lm:exact_CRBCeg} lead to 
\begin{eqnarray}
\mbox{CRB}^c_{eg}(\theta)  &\approx&   \frac{3\lambda^{2}}{2TD_{\mbox{SNR}}d^{2}\pi^{2}\cos^2(\theta)p_3(N)}\!\!\times\!\! \nonumber\\
           &&  \left   [p_2(N)  \!-\! 6(N \!- \!1)(6N^{2} \!-\! 15N \!+\! 11)\sin(\!\theta\!)\frac{d}{r} \right.   \label{eq:O_CRB_th} \\
           &&  \left.  +  \frac{1}{70}(2N - 1)(384N^{3} - 1353N^{2} + 1379N - 368)\frac{d^{2}}{r^{2}}\right.\nonumber\\
           &&  \left.  +  \frac{1}{14}(186N^{4}   -   1590N^{3} + 5351N^{2} - 6795N + 2890)\sin^{2}(\theta)\frac{d^{2}}{r^{2}}\right]  \nonumber\\
\mbox{CRB}^c_{eg}(r)  &\approx&  \frac{6r^{2}\lambda^{2}}{TD_{\mbox{SNR}}d^{4}\pi^{2} \cos^4(\theta)p_3(N)} \times \left[15 {r^{2}}-60(N-1)\sin(\theta) dr\right.\nonumber\\
            &&       \left.+ \frac{1}{14}   \left\{\sin^{2}(\theta)(1061N^{2}  -  2625N  +  2911) \right. \right.              \nonumber\\
            &&      \ \ \ \ \ \ \ \left.\left.+  225N^{2}  -  315N  -  135 \right\}d^2\right] \label{eq:O_CRB_r}
\end{eqnarray}
where $p_2(N)=(8N-11)(2N-1)$ and $p_3(N)=N(N^2-1)(N^2-4)$. 
\label{lm:OurTaylorExpansion}
\end{lemma}

These expressions are useful to compare our CRB to the one in \cite{elkorso} but also, they allow us to reveal how the different system parameters affect the localization performance. For example, one can see in particular that: 
\begin{itemize}
	\item The CRB decreases linearly w.r.t. the observation time as well as the deterministic SNR.
	\item The CRB goes to infinity when $\theta\rightarrow \pm \frac{\pi}{2}$ and the best localization results are obtained in the central direction, i.e., for $\theta=0$.
	\item Asymptotically (for large antenna sizes), the first term of the TE decreases in $N^3$, i.e., if we double the number of sensors, the CRB will be decreases approximately by a factor of $8$.
\end{itemize}
\subsubsection{Comparison with the {'CRB of the approximate model'}}
\label{comparison}
Let us first recall the CRB expressions given in \cite{elkorso} and based on the approximate model in (\ref{eq:approximated_tau}):
\begin{lemma}
The non-matrix expressions of the approximate conditional CRB \cite{elkorso} for a source in the near field, for $N\geq 3$ and $\theta\neq\pm\frac{\pi}{2}$, are given by
\begin{eqnarray}
\mbox{CRB}_{eg}(\theta)& = & \frac{3\lambda^2}{2TD_{\mbox{SNR}}d^2\pi^2\cos^2(\theta)p_3(N)} p_2(N)                                                                  \label{eq:K_CRB_th}\\
\mbox{CRB}_{eg}(r)     & = & \frac{6r^2\lambda^2}{TD_{\mbox{SNR}}d^4\pi^2\cos^4(\theta)p_3( N )}    \nonumber\\
                                      &    &  \left(15r^2  +  30dr(N  -1)\sin(\theta)  +  d^2p_2(N)\sin^2(\theta)\right) \label{eq:K_CRB_r}
\end{eqnarray}
\label{lm:lemma_korso}
\end{lemma}

Comparing lemma \ref{lm:OurTaylorExpansion} to lemma \ref{lm:lemma_korso}, one can make the following observations:
\begin{itemize}
  \item First, we note that the first and main term of equation (\ref{eq:O_CRB_th}) (resp. of equation (\ref{eq:O_CRB_r})) is equal to the first term of equation (\ref{eq:K_CRB_th}) (resp. of equation (\ref{eq:K_CRB_r})).
	\item We note also,  that the Taylor expansion (of the time delay) followed by CRB derivation leads to different results as compared to CRB derivation followed by Taylor expansion (of the CRB). The latter expansion being more accurate than the former as illustrated by our simulation results.
	\item The approxim.ate CRB of the angle estimate shown in lemma \ref{lm:lemma_korso} is independent of the range parameter (it is the same as the far field CRB)  while the expression  of $\mbox{CRB}(\theta)$ in lemma \ref{lm:OurTaylorExpansion} reveals how it is affected by the range parameter. In particular, at the first order, one can see that the CRB decreases (resp. increases) as a function of $\frac{d}{r}$ for $\theta\in[0,\frac{\pi}{2}[$ (resp. for $\theta\in]-\frac{\pi}{2},0]$).
\end{itemize}
\subsection{Exact unconditional CRB with EG}
\label{ExactCRBUeg} 
For unconditional CRB, the signal is assumed Gaussian complex circular with zero mean and variance $\sigma^{2}_{s}$. Assuming that $\boldsymbol{\xi} =  [\theta,r,\sigma_s^2,\sigma^2]^{T}$ is the vector of the unknown parameters, the log-likelihood function of the observed data is given by
$$\ln(\boldsymbol{\xi})=-NT \ln\pi - \ln det(\boldsymbol{\Sigma}) - T tr(\boldsymbol{\Sigma}^{-1} \hat{\bf R}) $$
where $\boldsymbol{\Sigma}=\sigma^{2}_{s} \boldsymbol{a}(\theta,r)\boldsymbol{a}(\theta,r)^{H} + \sigma^{2}{\bf I}_{N}$ (theoretical covariance matrix), $\hat{\bf R}=\frac{1}{T}\sum^{T}_{t=1} \boldsymbol{x}(t)\boldsymbol{x}^H(t)$ (sample estimate covariance matrix), tr(.) refers to the matrix trace and $(.)^{H}$ is the transpose conjugate operator. 

The Fisher Information matrix in this Gaussian case is given by
\begin{equation}
	[\mbox{FIM}(\boldsymbol{\boldsymbol{\xi}})]_{i,j} = T tr\left(\boldsymbol{\Sigma}^{-1} \frac{\partial \boldsymbol{\Sigma}}{\partial {\xi}_{i}} \boldsymbol{\Sigma}^{-1} \frac{\partial \boldsymbol{\Sigma}}{\partial {\xi}_{j}}\right)
	\label{eq:FIM_gaussian}
\end{equation}

Since the CRB is equal to the inverse of the Fisher information matrix and using the results of \cite{stoica} for the two desired localization parameters, we have the following lemma:
\begin{lemma} 
 The non-matrix expressions of the unconditional CRB in the equal gain case for a source in the near field, for $N \geq 3$ and $\theta \neq \pm\frac{\pi}{2}$ are given by
\begin{eqnarray}
\mbox{CRB}^{u}_{eg}(\theta)    &=&\left[\frac{(1+(\mbox{SNR})~~N)D_{\mbox{SNR}}}{(\mbox{SNR})^{2}~~N}\right]  \mbox{CRB}^{c}_{eg}(\theta)   \label{eq:exctCRBU_th}   \\
\mbox{CRB}^{u}_{eg}(r)         &=&\left[\frac{(1+(\mbox{SNR})~~N)D_{\mbox{SNR}}}{(\mbox{SNR})^{2}~~N}\right] \mbox{CRB}^{c}_{eg}(r)         \label{eq:exctCRBU_r}    \\
\mbox{CRB}^{u}_{eg}(r,\theta)  &=&\left[\frac{(1+(\mbox{SNR})~~N)D_{\mbox{SNR}}}{(\mbox{SNR})^{2}~~N}\right]  \mbox{CRB}^{c}_{eg}(r,\theta) \label{eq:exctCRBU_rth}
\end{eqnarray}
\label{lm:exact_CRBUeg}
where $\mbox{SNR}=\frac{\sigma^2_s}{\sigma^2}$ and $D_{\mbox{SNR}}\times \mbox{CRB}^{c}_{eg}$ represents the normalized $\mbox{CRB}^{c}_{eg}$ (see lemma \ref{lm:exact_CRBCeg}) depending on the localization parameters, the array geometry, and the sample size only.
\end{lemma}
\textbf{Proof}: The proof of this lemma can be deduced directly from lemma \ref{lm:exact_CRBUvg} proof given in appendix (\ref{App2}).

This result translates the fact that the unconditional CRB varies in a similar way as the conditional CRB w.r.t. variables $\theta$, $r$ and $T$.

However, concerning the SNR parameter and the number of sensors, it is interesting to observe that:
\begin{itemize}
	\item At low SNRs (i.e., if $\mbox{SNR}\times N << 1$), the CRB decreases quadratically (instead of linearly in the conditional case) w.r.t. the SNR and in $N^4$ (instead of $N^3$) in terms of the number of sensors.
	\item However, for large SNRs (i.e., if $\mbox{SNR}\times N>>1$), the unconditional CRB behaves similarly as the conditional CRB w.r.t. parameters SNR and $N$ (i.e., it decreases linearly with the SNR and in $N^3$ w.r.t. the number of sensors).
\end{itemize}

\section{Conditional and unconditional CRB derivation with Variable Gain (VG)}
\label{variable_gain}
In this case, we consider the same model as that described in Section \ref{data_model} except that the received power is variable from sensor to sensor. The $n^{th}$ output array is expressed as:
\begin{equation}
x_{n}(t)  = \gamma_{n}(r,\theta) s(t)e^{j\tau_{n}(r,\theta)} + v_{n}(t) = s(t)(\gamma_{n}(r,\theta) e^{j\tau_{n}(r,\theta)}) + v_{n}(t) \ \ \ t=1,\cdots,T  \label{eq:outputs_vg} 
\end{equation}
where $\gamma_{n}(r,\theta)$ represents the power profile of the $n^{th}$ sensor and is given by \cite{fresnel}
\begin{equation}
\gamma_{n}(r,\theta)= \frac{1}{d_{n} }= \frac{1}{r \sqrt{ 1 - \frac{2nd}{r} \sin \theta + \left(\frac{nd}{r}\right)^{2}}}
\label{eq:power_profile}
\end{equation}
$d_{n}$ being the distance between the source signal and the $n^{th}$ sensor.

As we can see, in this model both the time delay profile and the power profile carry information on the desired source location $(r,\theta)$. Our objective here, is to investigate the roles of both profiles in the performance limit given by the CRB.
\subsection{Exact conditional CRB with VG}
\label{CRBC_Vbl_gain}
Similarly to Section \ref{ExactCRBCeg}, the conditional CRB is given by equation (7) which reduces in the variable gain case to 
 \begin{equation}
  {\mbox{FIM}} = {\left[ \begin{array}{c  c }
                         \tilde{{\mathcal{\textbf{\textit{Q}}}}}&  {\bf 0} \\
                                   {\bf 0}&   \frac{NT}{\sigma^4}
                           \end{array} \right]}
 \label{eq:FIM_bd}
 \end{equation}
which translates the fact that the $\mbox{FIM}$ of the desired localization parameters is decoupled from the noise variance $\sigma^2$ but not from the source magnitude parameter as in the equal gain case. Hence the CRB matrix of the range and angle parameters is equal to the $2\times2$ top left sub-matrix of the following inverse matrix                      
\begin{displaymath}
	\tilde{{\mathcal{\textbf{\textit{Q}}}}}^{-1} = {\left[ \begin{array}{c  c | c   c}
																									f_{\theta \theta}        & f_{\theta r}          & \textbf{f}_{\theta \psi}  & \textbf{f}_{\theta \alpha} \\
																									f_{r \theta}             & f_{r r}               & \textbf{f}_{r \psi}       & \textbf{f}_{r \alpha}\\
																									\hline
																									\textbf{f}_{\psi \theta} & \textbf{f}_{\psi r}   & \textbf{F}_{\psi \psi}    & \textbf{F}_{\psi \alpha}\\
																									\textbf{f}_{\psi \alpha} & \textbf{f}_{\alpha r} & \textbf{F}_{\alpha \psi}  & \textbf{F}_{\alpha \alpha}\\
																									\end{array} \right]}^{-1}
\end{displaymath}
where  $ f_{r \theta} = f_{\theta r}$ and
\begin{eqnarray}
f_{\theta \theta} &=& 2T\tilde{D}_{\mbox{SNR}}(\left\| {\dot{\boldsymbol{\gamma}}}_{\theta} \right\|^{2} + (\boldsymbol{\gamma} \odot  \boldsymbol{\gamma})^T (\boldsymbol {\dot{\tau}}_{\theta}\odot \boldsymbol {\dot{\tau}}_{\theta}))  \label{ftt}\\
f_{r r}           &=& 2T\tilde{D}_{\mbox{SNR}}(\left\| {\dot{\boldsymbol{\gamma}}}_{r} \right\|^{2} + (\boldsymbol{\gamma} \odot  \boldsymbol{\gamma})^T (\boldsymbol {\dot{\tau}}_{r}\odot \boldsymbol {\dot{\tau}}_{r}))  \label{frr}\\
f_{r \theta}      &=& 2T\tilde{D}_{\mbox{SNR}}(({\dot{\boldsymbol{\gamma}}}^{H}_{\theta}{\dot{\boldsymbol{\gamma}}}_{r})  + (\boldsymbol{\gamma} \odot  \boldsymbol{\gamma})^T (\boldsymbol {\dot{\tau}}_{\theta}\odot \boldsymbol {\dot{\tau}}_{r})) \label{frt}                  
\end{eqnarray}

Note that $\tilde{D}_{\mbox{SNR}}$ is defined here similarly to ${D}_{\mbox{SNR}}$ ($\tilde{D}_{\mbox{SNR}}=\frac{\left\|{\boldsymbol \alpha}\right\|^2}{T\sigma^2}$) but it is implicitly scaled by the square of the range value which has been extracted from the received signal amplitude in (\ref{eq:outputs_vg}) in order to explicit the role of the power profile.

Consequently, $\tilde{D}_{\mbox{SNR}}$ unit is square meter (while ${D}_{\mbox{SNR}}$ is a constant without unit). Also, this way of normalizing the signals leads to CRB expressions in the variable gain case that are scaled (multiplied) by a factor of $r^2$ as compared to the equal gain case\footnote{This can be seen as if in the equal gain case, the received power is set equal to a constant value while in the variable gain case, it is the transmit power that is set equal to a constant.}. For this reason, when comparing the equal gain scenario to the variable gain one in Section \ref{Simulation}, we normalize the received signal power by dividing $\left\|\boldsymbol{\alpha}\right\|^2$ by $r^2$ in the former case.

Moreover, the vectors of size $T\times1$, $\textbf {f}_{\psi \theta}$, $\textbf {f}_{\theta \psi}^{T}$, $\textbf {f}_{\psi r}$ and $\textbf {f}_{r \psi}^{T}$ are given by 
\begin{eqnarray*}
\textbf {f}_{\psi \theta}   &=& \textbf {f}_{\theta \psi}^{T}=\frac{2}{\sigma^{2}} (\boldsymbol{\gamma} \odot  \boldsymbol{\gamma})^T \boldsymbol {\dot{\tau}}_{\theta} \\
\textbf {f}_{\psi r}        &=& \textbf {f}_{r \psi}^{T}=\frac{2}{\sigma^{2}} (\boldsymbol{\gamma} \odot  \boldsymbol{\gamma})^T \boldsymbol {\dot{\tau}}_{r}   \\
\textbf {f}_{\alpha \theta} &=& \textbf {f}_{\theta \alpha}^{T} =\frac{2}{\sigma^{2}}(\boldsymbol{\gamma}^{T}{\dot{\boldsymbol{\gamma}}}_{\theta})\boldsymbol{\alpha}\\
\textbf {f}_{\alpha r}      &=& \textbf {f}_{r \alpha}^{T} =\frac{2}{\sigma^{2}}(\boldsymbol{\gamma}^{T}{\dot{\boldsymbol{\gamma}}}_{r}) \boldsymbol{\alpha}
\end{eqnarray*}
where 
\begin{eqnarray*}
\boldsymbol{\gamma}                &=& \left[\gamma_0 (r,\theta),...,\gamma_{N-1}(r,\theta)\right]^T  \\
\dot{\boldsymbol{\gamma}}_{\theta} &=& \left[\frac{\partial{\gamma_{0}}}{\partial{\theta}},...,\frac{\partial{\gamma_{N-1}}}{\partial{\theta}}\right]^T  \\
\dot{\boldsymbol{\gamma}}_{r}      &=& \left[\frac{\partial{\gamma_{0}}}{\partial{r}},...,\frac{\partial{\gamma_{N-1}}}{\partial{r}}\right]^T
\end{eqnarray*}
Finally, matrices \textbf {F}$_{\psi \psi}$, \textbf {F}$_{\alpha \alpha}$, \textbf {F}$_{\psi \alpha}$ of size $T \times T$, are given by
\begin{equation*}
\textbf {F}_{\psi \psi} = \frac{2}{\sigma^{2}} \left\|\boldsymbol{\gamma} \right\|^{2} \mbox{diag}(\boldsymbol{\alpha} \odot  \boldsymbol{\alpha}),~~ \textbf {F}_{\psi \psi}=\frac{2}{\sigma^{2}} \left\|\boldsymbol{\gamma} \right\|^{2} {\bf I}_T ~~ \mbox{and} ~~ \textbf {F}_{\psi \alpha}=\boldsymbol{0}.
\end{equation*}

Again by using the Schur's matrix inversion lemma, one can obtain:
\begin{lemma}
 The non-matrix expressions of the conditional CRB in the variable gain case  for a source in the near field, for $N \geq 3$ and $\theta \neq \pm\frac{\pi}{2}$, are given by
\begin{eqnarray}
\mbox{CRB}^{c}_{vg}(r)        =\left(\frac{1}{2T\tilde{D}_{\mbox{SNR}}}\right)\frac{E_{vg}(\theta)}{E_{vg}(\theta)E_{vg}(r)-E_{vg}(r,\theta)^{2}}      &&\label{eq:exctCRBCvg_r} \\
\mbox{CRB}^{c}_{vg}(\theta)   =\left(\frac{1}{2T\tilde{D}_{\mbox{SNR}}}\right)\frac{E_{vg}(r)}{E_{vg}(\theta)E_{vg}(r)-E_{vg}(r,\theta)^{2}}           &&\label{eq:exctCRBCvg_th} \\
\mbox{CRB}^{c}_{vg}(r,\theta) =\left(\frac{1}{2T\tilde{D}_{\mbox{SNR}}}\right)\frac{E_{vg}(r,\theta)}{E_{vg}(\theta)E_{vg}(r)-E_{vg}(r,\theta)^{2}}    &&\label{eq:exctCRBCvg_rth}
\end{eqnarray}
where 
\begin{eqnarray}
E_{vg}(\theta)   & = & \left\|\dot{\boldsymbol{\gamma}}_{\theta}\right\|^{2}+ (\boldsymbol{\gamma} \odot  \boldsymbol{\gamma})^T (\boldsymbol {\dot{\tau}}_{\theta}\odot \boldsymbol {\dot{\tau}}_{\theta})   -  \frac{1}{\left\|\boldsymbol{\gamma}\right\|^{2}}\left[((\boldsymbol{\gamma} \odot  \boldsymbol{\gamma})^T \boldsymbol {\dot{\tau}}_{\theta})^{2}+ (\boldsymbol{\gamma}^{T}\dot{\boldsymbol{\gamma}}_{\theta})^{2}\right] \label{eq:Evg_th}\\
E_{vg}(r)        & = & \left\|\dot{\boldsymbol{\gamma}}_{r}\right\|^{2}+ (\boldsymbol{\gamma} \odot  \boldsymbol{\gamma})^T (\boldsymbol {\dot{\tau}}_{r}\odot \boldsymbol {\dot{\tau}}_{r})    - \frac{1}{\left\|\boldsymbol{\gamma}\right\|^{2}}\left[((\boldsymbol{\gamma} \odot  \boldsymbol{\gamma})^T \boldsymbol {\dot{\tau}}_{r})^{2}+ (\boldsymbol{\gamma}^{T}\dot{\boldsymbol{\gamma}}_{r})^{2}\right] \label{eq:Evg_r}\\
E_{vg}(r,\theta) & = & \dot{\boldsymbol{\gamma}}^{T}_{\theta}\dot{\boldsymbol{\gamma}}_{r} + (\boldsymbol{\gamma} \odot  \boldsymbol{\gamma})^T (\boldsymbol {\dot{\tau}}_{\theta}\odot \boldsymbol {\dot{\tau}}_{r})  -  \frac{1}{\left\|\boldsymbol{\gamma}\right\|^{2}}\left[(\boldsymbol{\gamma} \odot  \boldsymbol{\gamma})^T \boldsymbol {\dot{\tau}}_{\theta}(\boldsymbol{\gamma} \odot  \boldsymbol{\gamma})^T \boldsymbol {\dot{\tau}}_{r} + (\boldsymbol{\gamma}^{T}\dot{\boldsymbol{\gamma}}_{r})(\boldsymbol{\gamma}^{T}\dot{\boldsymbol{\gamma}}_{\theta})\right] \label{eq:Evg_rth}
\end{eqnarray}
\label{lm:exact_CRBCvg}
\end{lemma}
\textbf{Proof}: See appendix \ref{App1}

One can observe, that the equal gain case (i.e., lemma \ref{lm:exact_CRBCeg}) is a particular situation of lemma \ref{lm:exact_CRBCvg} where $\boldsymbol{\gamma}={\bf 1}_N$ and $D_{\mbox{SNR}}$ is divided by $r^2$.
\subsection{Exact unconditional CRB with VG}
Under the stochastic Gaussian model of Section \ref{ExactCRBUeg}, the unconditional CRB with variable gain can be expressed as stated by the following lemma:
\begin{lemma}
The non-matrix expressions of the exact unconditional CRB in the variable case for a source in the near field, for $N \geq 3$ and $\theta \neq \pm\frac{\pi}{2}$, are given by
\begin{center}
\begin{tabular}{|c c|}
\hline
$\mbox{CRB}^{u}_{vg}(\theta)   = \left(\frac{(1+\tilde{\mbox{SNR}} \left\|\boldsymbol{\gamma}\right\|^{2})\tilde{D}_{\mbox{SNR}}}{(\tilde{\mbox{SNR}})^{2}\left\|\boldsymbol{\gamma}\right\|^{2}}\right)\mbox{CRB}^{c}_{vg}(\theta) $ &  \stepcounter{equation}\thetag{\theequation} \label{eq:exctCRBUvg_th}\\%
$\mbox{CRB}^{u}_{vg}(r)        =\left(\frac{(1+\tilde{\mbox{SNR}} \left\|\boldsymbol{\gamma}\right\|^{2})\tilde{D}_{\mbox{SNR}}}{(\tilde{\mbox{SNR}})^{2}\left\|\boldsymbol{\gamma}\right\|^{2}}\right)\mbox{CRB}^{c}_{vg}(r)$&  \stepcounter{equation}\thetag{\theequation}\label{eq:exctCRBUvg_r}\\
$\mbox{CRB}^{u}_{vg}(r,\theta) =\left(\frac{(1+\tilde{\mbox{SNR}} \left\|\boldsymbol{\gamma}\right\|^{2})\tilde{D}_{\mbox{SNR}}}{(\tilde{\mbox{SNR}})^{2}\left\|\boldsymbol{\gamma}\right\|^{2}}\right)\mbox{CRB}^{c}_{vg}(r,\theta)$&  \stepcounter{equation}\thetag{\theequation}\label{eq:exctCRBvg_rth}\\
\hline
\end{tabular}
\end{center}
%
%
\label{lm:exact_CRBUvg}
\end{lemma}
\textbf{Proof}: See appendix \ref{App2}

Similarly to the equal gain case, $\tilde{D}_{\mbox{SNR}}\times \mbox{CRB}^{c}_{eg}$ represents a normalized CRB independent from the SNR value. Also, $\tilde{\mbox{SNR}}$ is defined in the same way as the SNR   in lemma \ref{lm:exact_CRBUeg} except that it has been implicitly scaled by the square of the range parameter ($\tilde{\mbox{SNR}}$ unit is now square meter).
\subsection{Comparison between VG and EG cases}
\label{Vble_cst_gains}
By considering the extra-information given by the received power profile, one is able to achieve better localization performance as will be shown later in Section \ref{Simulation}. To better compare the CRB expressions in the constant and variable gain cases, we provide here the first term of their Taylor expansion
\begin{eqnarray*}
\mbox{CRB}^c_{eg}(\theta)&\approx& \frac{3 \lambda^{2}p_2(N)}{2\pi^{2}TD_{\mbox{SNR}}d^{2}\cos^2(\theta)p_3(N)} \\
\mbox{CRB}^c_{vg}(\theta)&\approx& \frac{3 \lambda^{2}r^2p_2(N)\left(1+\lambda^2f_1(\theta)\right)}{2\pi^{2}T\tilde{D}_{\mbox{SNR}}d^2\cos^2(\theta)p_3(N)\left(1+\lambda^2f_2(\theta)\right)}\\
\mbox{CRB}^c_{eg}(r)     &\approx& \frac{90\lambda^{2}r^4}{\pi^{2}TD_{\mbox{SNR}}d^4\cos^4(\theta)p_3(N)}\\
\mbox{CRB}^c_{vg}(r)     &\approx& \frac{90\lambda^{2}r^6}{\pi^{2}T\tilde{D}_{\mbox{SNR}}d^4\cos^4(\theta)p_3(N)\left(1+\lambda^2f_2(\theta)\right)}
\end{eqnarray*}
where $f_1(\theta)=\frac{15\sin^2(\theta)}{\pi^2d^2\cos^4(\theta)p_2(N)}$ and $f_2(\theta)=\frac{15\sin^2(\theta)}{\pi^{2}d^2\cos^4(\theta)(N^2-4)}$. 
From these expressions, one can see that the two CRBs are quite similar for sources located in the central direction (i.e., for small $\theta$ values) while at lateral directions (i.e., $\left|\theta\right|$ close to $\frac{\pi}{2}$) the variable gain CRB is much lower than the equal gain one. This translates the fact that when the source location information contained in the time delay profile becomes weak, it is somehow partially compensated by the location information contained in the received power profile especially for the estimation of the range value. This observation is illustrated by the simulation experiments (cf. Fig. \ref{fig:Cst_Vble_Gains1}, Fig. \ref{fig:Cst_Vble_Gains2} and Fig. \ref{fig:Cst_Vble_Gains3}) given in Section \ref{Simulation}.

\section{Near field localization region (NFLR)}
\label{near_field}
The radiating near field or Fresnel region is the region between the near and far fields \cite{fresnel}  corresponding to the space region defined by equation (\ref{eq:Fresnel_region}). The latter depends on the source-antenna range, the signal wavelength, and the antenna aperture.

Note that the lower bound of Fresnel region is related to the fact that in the immediate vicinity of the antenna, the fields are predominately reactive fields meaning that the E and H fields are orthogonal \cite{fresnel}. Therefore the space region given by $r\leq 0.62\left(\frac{d^3(N-1)^3}{\lambda}\right)^{\frac{1}{2}}$ should be kept out of the localization region. However, the upper bound of the Fresnel region as given in equation (\ref{eq:Fresnel_region}) does not take into consideration the localization performance limit. For this reason, we suggest to define the near field localization region (NFLR)  based on a target estimation performance relative to the application at hand. More precisely, if $Std_{max}>0$ is the maximum standard deviation of the localization error that is tolerated by the considered application, i.e., 
\begin{equation}
\sqrt{\E\left(\left\|\hat{\bf d}-{\bf d}\right\|^2\right)}\leq Std_{max}
\label{eq:condition_eps}
\end{equation}
where ${\bf d}=(x,y)^T$ (resp. $\hat{{\bf d}}$) refers to the location vector (resp. its estimate), then the NFL region can be defined as the one for which the minimum standard deviation (given by the square root of the CRB) satisfies condition (\ref{eq:condition_eps}).

Since $\E\left(\left\|\hat{\bf d}-{\bf d}\right\|^2\right)=\E((\hat{x}-x)^2)+\E((\hat{y}-y)^2)$, the previous condition on the minimum MSE can be expressed as 
\begin{equation}
\sqrt{\mbox{CRB}(x)+\mbox{CRB}(y)}\leq Std_{max}
\label{eq:condition_CRB}
\end{equation}

Now, the source coordinates can be rewritten according to 
\begin{eqnarray}
x&=&r\sin(\theta) = g_x(r,\theta) \nonumber\\
y&=&r\cos(\theta) = g_y(r,\theta) \nonumber
\end{eqnarray}
and hence, by using the delta method in \cite{delta_method}, one can express
\begin{equation}
\mbox{CRB}(x)+\mbox{CRB}(y) = \nabla g_x^T(r,\theta)\C \nabla g_x(r,\theta)  + \nabla g_y^T(r,\theta)\C \nabla g_y(r,\theta)
\label{eq:CRB_x_y}
\end{equation}
where  
\begin{eqnarray*}
\nabla g_x(r,\theta) & = & \left[\frac{\partial g_x}{\partial r} \ \ \frac{\partial g_x}{\partial \theta}\right]^T = \left[\cos(\theta)  \ \ \ \-r\sin(\theta)\right]^T \\
\nabla g_y(r,\theta) & = & \left[\frac{\partial g_y}{\partial r} \ \ \frac{\partial g_y}{\partial \theta}\right]^T = \left[\sin(\theta)  \ \ \ \ r\cos(\theta)\right]^T    \\
\C                   & = &\left[\begin{array}{cc} \mbox{CRB}(\theta) & \mbox{CRB}(r,\theta) \\ \mbox{CRB}(\theta,r) & \mbox{CRB}(r) \end{array}\right]
\end{eqnarray*}

A straightforward derivation of (\ref{eq:CRB_x_y}) leads to 
\begin{equation}
\mbox{CRB}(x)+\mbox{CRB}(y) = r^2\mbox{CRB}(\theta)+\mbox{CRB}(r)
\label{eq:CRB_x_y_1}
\end{equation}
and therefore the NFL region is defined as the one corresponding to 
\begin{equation}
\sqrt{r^2\mbox{CRB}(\theta)+\mbox{CRB}(r)}\leq Std_{max}
\label{eq:CRB_x_y_2}
\end{equation}

An alternative 	approach would be to use a maximum tolerance value on the relative location error, i.e., a threshold value $\epsilon$ such that 
\begin{equation}
\sqrt{\frac{\E\left(\left\|\hat{\bf d}-{\bf d}\right\|^2\right)}{\left\|{\bf d}\right\|^2}}\leq \epsilon
\label{eq:E_eps}
\end{equation}
which corresponds to 
\begin{equation}
\sqrt{\mbox{CRB}(\theta) + \frac{\mbox{CRB}(r)}{r^2}}\leq \epsilon
\label{eq:CRB_r_theta_simple}
\end{equation}

For example, in the conditional case, equation (\ref{eq:CRB_r_theta_simple}) becomes
\begin{equation}
\frac{1}{2T\tilde{D}_{\mbox{SNR}}}\G_N(r,\theta)\leq \epsilon^2
\label{eq:35_transfrmee}
\end{equation}
where $$\G_N(r,\theta) = \frac{E_{vg}(\theta)+{E_{vg}(r)}/{r^2}}{E_{vg}(\theta)E_{vg}(r)-E_{vg}(r,\theta)^2}$$
From a practical point of view, equation (\ref{eq:35_transfrmee}) can be used to tune the system parameters in order to achieve a desired localization performance. Different scenarii can be considered, according to the parameter, we can (or wish to) tune. \\[0.5cm]
{\bf Scenario 1:} One can define the minimum observation time to achieve a desired localization performance at a given location and a given SNR value as
\begin{equation}
T_{min}(r,\theta)=\frac{\G_N(r,\theta)}{2\epsilon^2\tilde{D}_{\mbox{SNR}}}
\label{eq:T_min}
\end{equation}
Similarly, one can also define the minimum SNR value for a target localization quality as 
\begin{equation}
\tilde{D}_{\mbox{SNR}_{min}}(r,\theta)=\frac{\G_N(r,\theta)}{2\epsilon^2T}
\label{eq:DSNR_min}
\end{equation}

In Section \ref{Simulation}, we provide simulation examples to illustrate the variation of these two parameters w.r.t. the source location. \\[0.5cm]
{\bf Scenario 2:} The previous parameters can be also be defined for a desired localization region $R_d$ (instead of a single location point $(r,\theta)$) as
\begin{eqnarray}
T_{min}(R_d)       &=& \max_{(r,\theta)\in R_d} T_{min}(r,\theta)        \label{eq:Tmin_Rd} \\
D_{\mbox{SNR}_{min}}(R_d) &=& \max_{(r,\theta)\in R_d} D_{\mbox{SNR}_{min}}(r,\theta)  \label{eq:DSNR_Rd}
\end{eqnarray}

For example, if we are interested into the surveillance of a space sector limited by $r_{min}<r<r_{max}$ and $-\theta_{max}<\theta<\theta_{max}$, then 
\begin{eqnarray}
T_{min} & =       & \frac{1}{2\epsilon^2\tilde{D}_{\mbox{SNR}}}\max_{R_d}\G_N(r,\theta) \nonumber\\
        & \approx & \frac{1}{2\epsilon^2\tilde{D}_{\mbox{SNR}}}\G_N(r_{max},\theta_{max}) \label{eq:T_min_equ2} \\
\tilde{D}_{\mbox{SNR}_{min}} & = 			& \frac{1}{2\epsilon^2T}\max_{R_d}\G_N(r,\theta) \nonumber\\
										& \approx & \frac{1}{2\epsilon^2T}\G_N(r_{max},\theta_{max}) \label{eq:SNRT_min_equ2}
\end{eqnarray}
where the second equality holds from the observation that, away from the origin, $\G_N$ is a decreasing function w.r.t. the angular and range parameters\footnote{This is not an exact and proven statement but just an approximation that expresses the fact that the localization accuracy decreases when the source moves away from the antenna or towards its lateral directions.}. \\[0.5cm]
{\bf Scenario 3:} One can also wish to optimize the number of sensors with respect to a desired localization region $R_d$ and for a target localization quality $\epsilon$. In that case, the minimum number of sensors needed to achieve the target quality can be calculated as 
\begin{equation*}
N_{min}=\arg\min_{N} \left\{N\in \mathbb{N}^* \left|\G_N(r,\theta)\leq2\epsilon^2T\tilde{D}_{\mbox{SNR}} \ \ \forall (r,\theta)\in R_d\right.\right\}
\label{eq:Nmin}
\end{equation*}

\section{Simulation}
\label{Simulation}
In this Section, three experimental sets are considered. The first one is to compare the provided exact CRB expressions of lemma \ref{lm:exact_CRBCeg} with the CRB expressions given in \cite{elkorso}. Also, in that first experiment, we  used the maximum likelihood approach (with the exact model in (\ref{eq:outputs_eg}) and (\ref{eq:exact_tau})) to validate our exact CRB derivation and to illustrate the gain in location estimation accuracy when using (\ref{eq:exact_tau}) instead of (\ref{eq:approximated_tau}) in the data model. In the second experiment, we investigate the effect of considering the variable gain model instead of the constant gain model for near field source localization. Finally, the third experiment is to illustrate the usefulness of the near field localization region as compared to the standard Fresnel region.

In all our simulations, we consider a uniform linear antenna with $N=15$ sensors and iter-element spacing $d=\frac{\lambda}{2}$ $(\lambda=0.5m)$ receiving signals from one near field source located at $(r,\theta)$. The sample size is $T=90$ (unless stated otherwise) and the observed signal is corrupted by a white  Gaussian circular noise of variance $\sigma^2$. In the conditional case, the source signal is of unit amplitude (i.e., $\alpha(t)=1\ \ \forall t$).

The dotted vertical plots in all figures represent the upper and lower range limits of the Fresnel region given by (\ref{eq:Fresnel_region}). 

\subsection{Experiment 1: Comparison with existing work}
\label{Experiment1}
\begin{figure}%
\centerline{\includegraphics[width=9cm]  {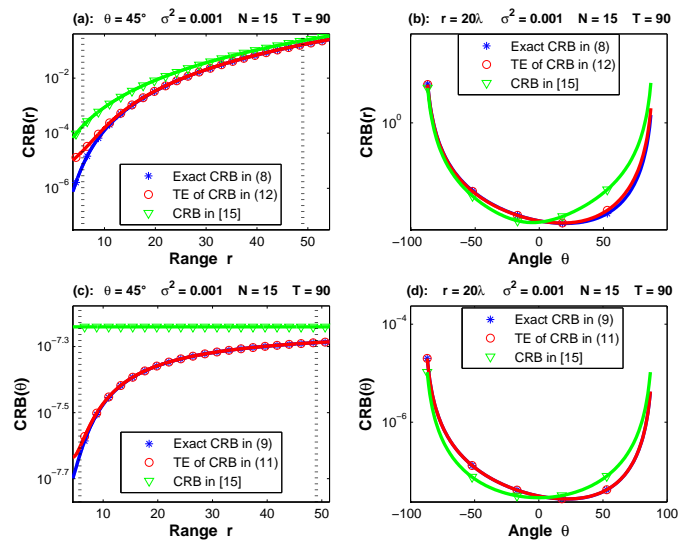}}%
\caption{CRB comparison: Exact conditional CRB versus approximate CRB in \cite{elkorso} in high SNR case}%
\label{fig:Comp_korso1}%
\end{figure}

\begin{figure}
\centerline{\includegraphics[width=9cm]  {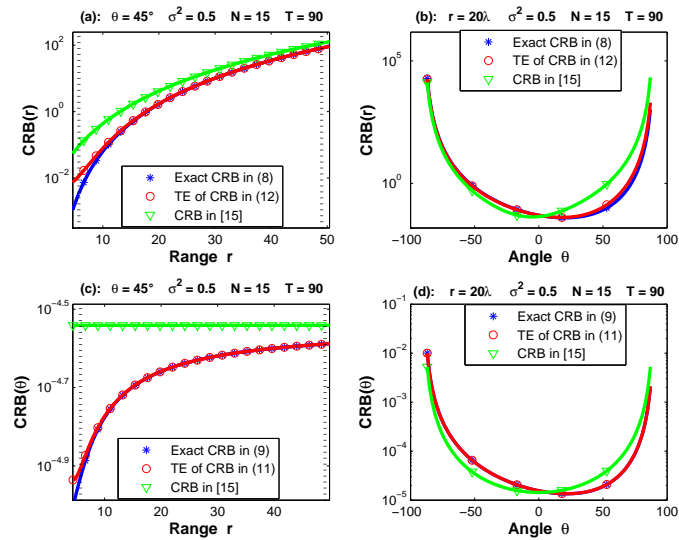}}%
\caption{CRB comparison: Exact conditional CRB versus approximate CRB in \cite{elkorso}  in low SNR case}%
\label{fig:Comp_korso2}%
\end{figure}

In Fig. \ref{fig:Comp_korso1}, we compare the three CRB expressions for the source location parameter estimates  versus the range values in the interval $\left[0\ \ 50m\right]$ and versus angle values\footnote{Note that the  curves w.r.t. the angle parameter (i.e., Fig. \ref{fig:Comp_korso1}, Fig. \ref{fig:Comp_korso2}, Fig. \ref{fig:Cst_Vble_Gains3}, Fig. \ref{fig:NFL_5} and Fig. \ref{fig:NFL_7}) are not symmetrical around $\theta=0$ because we have chosen the first sensor for the time reference  as shown in Fig. \ref{fig:model}.} in the interval $[-90^o\ \ 90^o]$. The noise level is set to $\sigma^2=0.001$ which corresponds to $\mbox{SNR} = 30\ dB$. A similar comparison leading to similar results is given in Fig. \ref{fig:Comp_korso2}  for a noise level  set equal to $\sigma^2 = 0.5$ which corresponds to $\mbox{SNR} \approx 3\ dB$. The source angle is $\theta = 45^o$ in the comparison versus range values and $r = 20\lambda$ in the comparison versus angle values.

From these figures, one can observe a non negligeable difference between the exact CRB and the proposed one in \cite{elkorso} especially at low range values: i.e., the given CRB in \cite{elkorso} can be up to $30$ times larger than the exact CRB. Also, from Fig. \ref{fig:Comp_korso1}.(c) and Fig. \ref{fig:Comp_korso2}.(c), contrary to the given CRB in \cite{elkorso}, the exact one varies with the range value with a relative difference varying  from approximately $60\%$ for small ranges to 0 when $r$ goes to infinity. 

From Fig. \ref{fig:Comp_korso1}.(b) - Fig. \ref{fig:Comp_korso2}.(b), one can observe that the lowest CRB is obtained in the central directions, this observation can be seen from the TE given in lemma \ref{lm:OurTaylorExpansion} where the factor $\frac{1}{\cos(\theta)}$ is minimum for this directions and goes to infinity when $|\theta| \rightarrow \frac{\pi}{2}$.

Note that the provided Taylor expansion of the exact CRB is more accurate than the one obtained by expanding the time delay expression before CRB derivation i.e., the one in \cite{elkorso}.

\begin{figure}
\centerline{\includegraphics[width=9cm]  {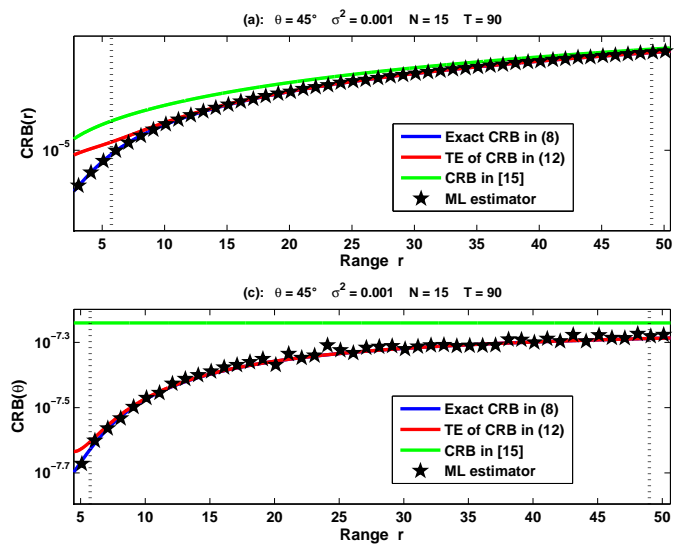}}%
\caption{Exact conditional CRB validation}%
\label{fig:CRB_validation}%
\end{figure}

To validate our exact CRB expression (see Fig. \ref{fig:CRB_validation}), we have considered the Maximum likelihood (ML) algorithm. The ML function is computed for the equal gain conditional model, and its maximization is ensured by using Newton-Raphson method (6000 Monte Carlo runs are considered for the ML estimator MSE plot in Fig. \ref{fig:CRB_validation}). One can see that these CRB expressions can be reached and that we can gain up to $60\%$ of MSE reduction by avoiding the standard model approximation in (\ref{eq:approximated_tau}).

\subsection{Experiment 2: EG versus VG cases}
\label{Experiment2}
\begin{figure}%
\centerline{\includegraphics[width=9cm]  {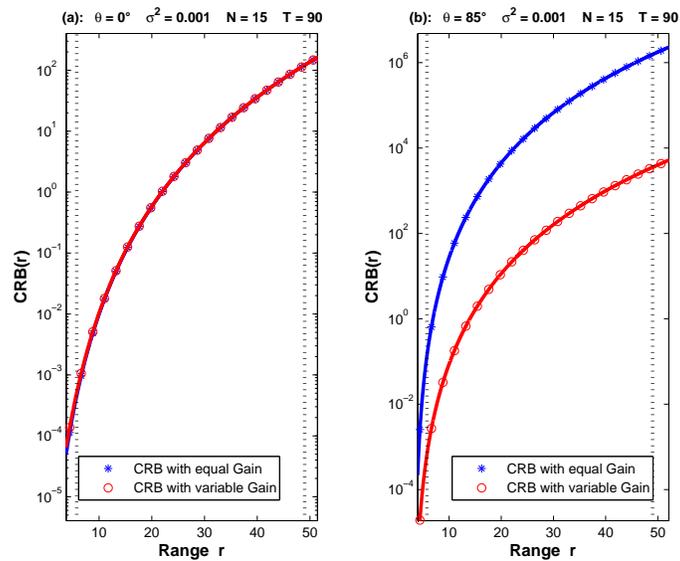}}%
\caption{CRB comparison: Equal Gain versus Variable Gain cases for range estimation}%
\label{fig:Cst_Vble_Gains1}%
\end{figure}

\begin{figure}%
\centerline{\includegraphics[width=9cm]  {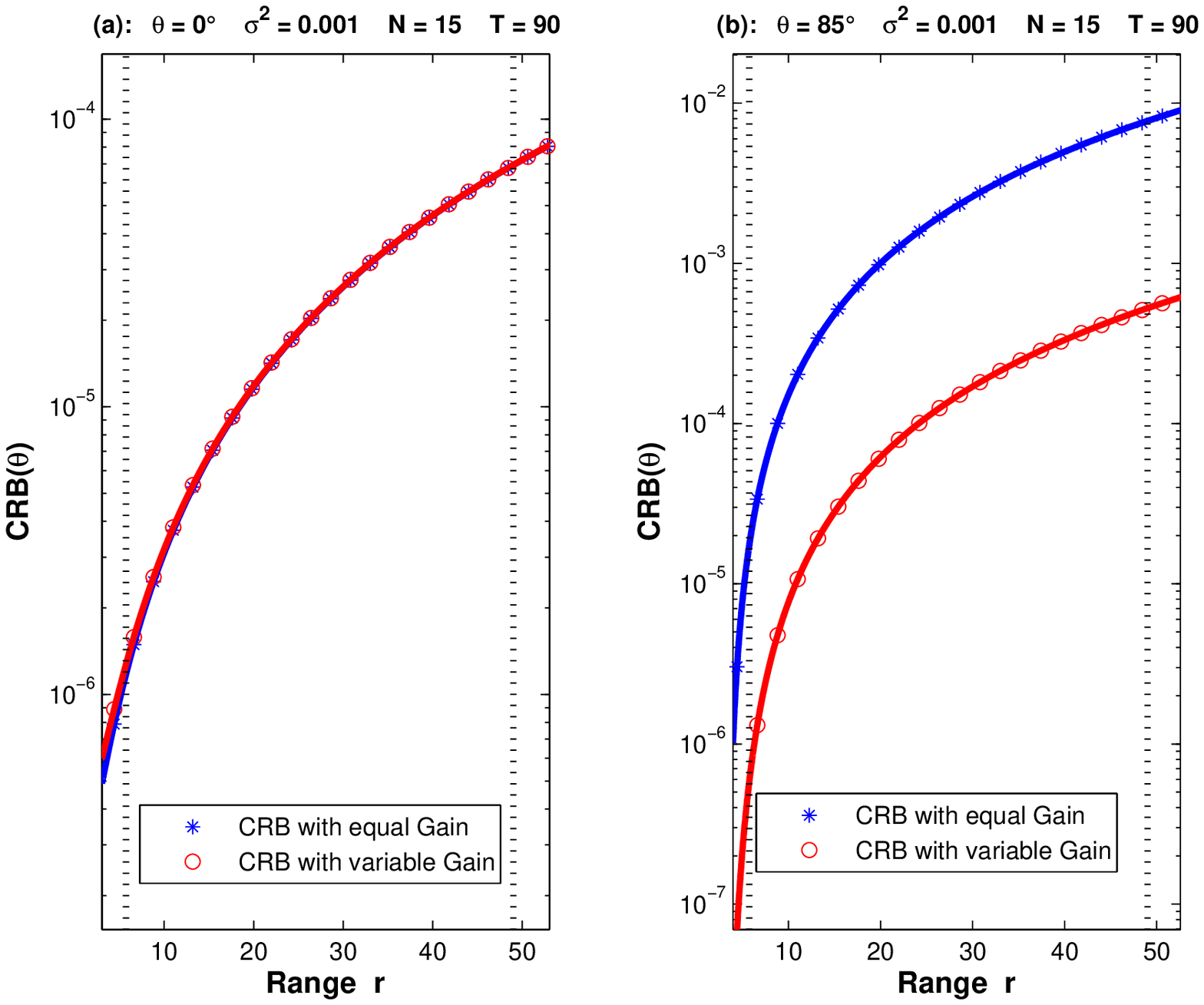}}%
\caption{CRB comparison: Equal Gain versus Variable Gain cases for angle estimation}%
\label{fig:Cst_Vble_Gains2}%
\end{figure}

To this end, we have to ensure first that the received power in the two cases (i.e., constant and variable gain cases) is the same for the reference sensor (i.e., dividing the power of constant gain case per the square of the range as explained in Section \ref{CRBC_Vbl_gain}). To better compare CRB expressions, we consider two contexts where $\theta=0^o$  for the first context (central direction) and $\theta=85^o$  for the second context (lateral direction). One can observe from Fig. \ref{fig:Cst_Vble_Gains1} and Fig. \ref{fig:Cst_Vble_Gains2} that, for small $\left|\theta\right|$ values, the constant gain CRB is quite similar to the variable gain one while at lateral direction (i.e., $\left|\theta\right|$ close to $\frac{\pi}{2}$) the variable gain CRB is much lower than the equal gain one, due to the extra information brought by the considered gain profile. 

\begin{figure}%
\centerline{\includegraphics[width=9cm]  {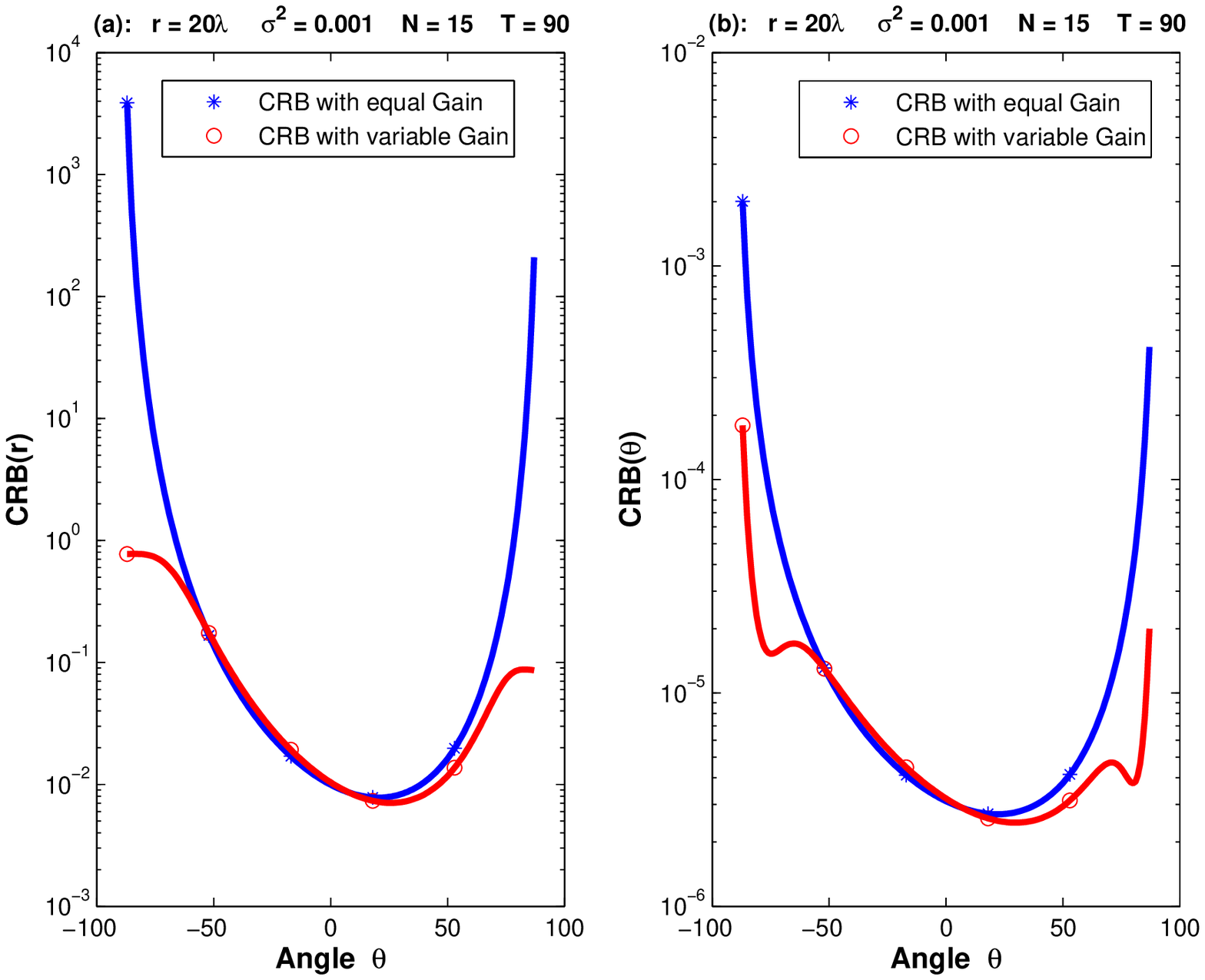}}%
\caption{CRB comparison of the Equal Gain and Variable Gain cases versus angle value}%
\label{fig:Cst_Vble_Gains3}%
\end{figure}

This can be seen again from Fig. \ref{fig:Cst_Vble_Gains3} where we can observe the large  CRB difference for high $\left|\theta\right|$ values.

\subsection{Experiment 3: Near Field Localization Region}
\label{Experiment3}
\begin{figure}%
\centerline{\includegraphics[width=9cm]  {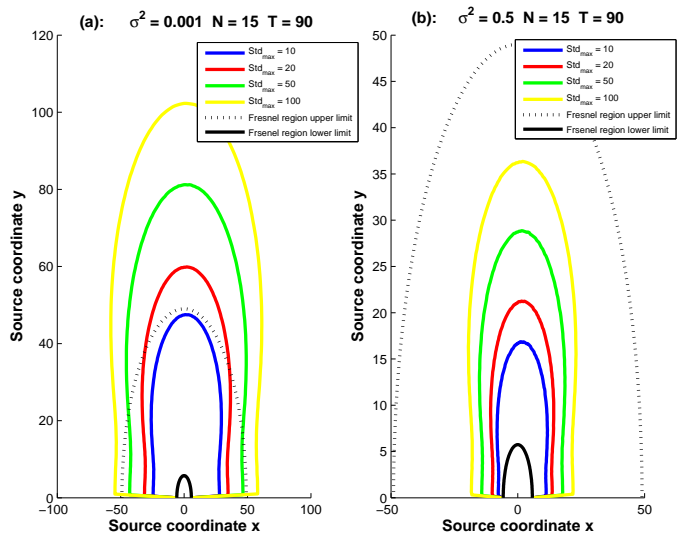}}%
\caption{Near field localization regions for different values of the target quality: $Std_{max}$}%
\label{fig:NFL_1}%
\end{figure}

The plots in Fig. \ref{fig:NFL_1} represent the upper limit of the NFL region for different tolerance values. From this figure, one can observe that the Fresnel region is not appropriate  to characterize the localization performance. Indeed, depending on the target quality, one can have space locations (i.e., sub-regions) in the Fresnel region that are out of the NFLR. Inversely, we have space locations not part of the Fresnel region that are attainable, i.e., they belong to the NFLR.

\begin{figure}%
\centerline{\includegraphics[width=9cm]  {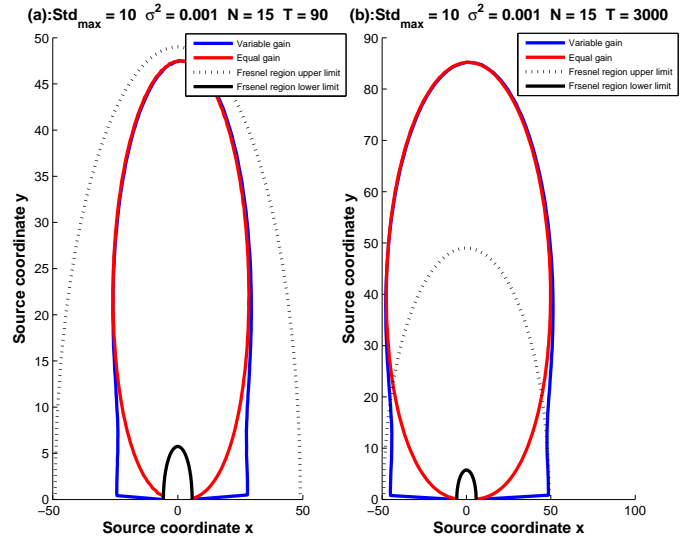}}%
\caption{Comparison of NFL regions in equal and variable gain cases}%
\label{fig:NFL_2}%
\end{figure}

Fig. \ref{fig:NFL_2} compares the NFL region in the variable gain and equal gain cases with $\mbox{SNR} = 30\ dB$. One can observe that in the lateral directions the NFLR associated to the variable gain model is much larger than its counterpart associated to the standard equal gain model. Also, in the short observation time context (i.e., Fig. \ref{fig:NFL_2}.(a)) the NFLR is included in the Fresnel region while for large observation time (i.e., Fig. \ref{fig:NFL_2}.(b)) the NFLR region is much more expanded and contains most of the Fresnel region\footnote{Except for the extreme lateral directions where the target quality can never be met since the CRB goes to infinity for  $|\theta| \rightarrow \frac{\pi}{2}$.}.

\begin{figure}%
\centerline{\includegraphics[width=9cm]  {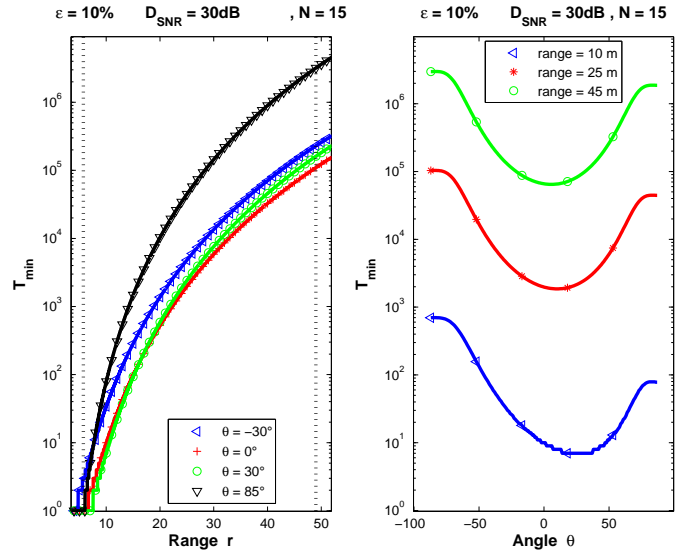}}%
\caption{Variation of the minimum observation time versus the source location parameters}%
\label{fig:NFL_5}%
\end{figure}

\begin{figure}%
\centerline{\includegraphics[width=9cm]  {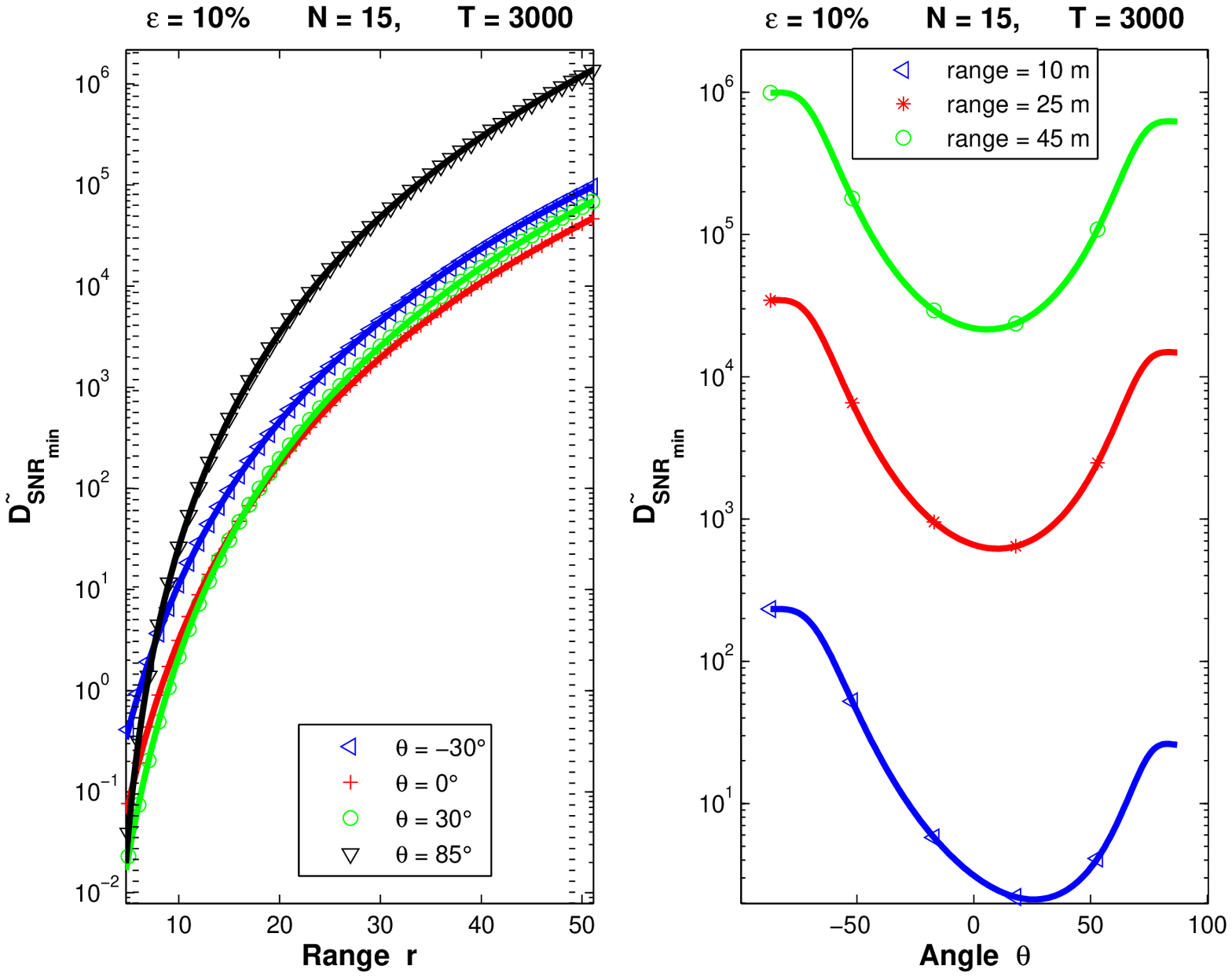}}%
\caption{Variation of the minimum deterministic SNR versus the source location parameters}%
\label{fig:NFL_7}%
\end{figure}

In Fig. \ref{fig:NFL_5} - Fig. \ref{fig:NFL_7}, we illustrate the variation of the two parameters $T_{min}$ and $\tilde{D}_{\mbox{SNR}_{min}}$ w.r.t. the source location parameters and for a relative tolerance error equal to $\epsilon = 10\%$. From these figures, one can observe that $\tilde{D}_{\mbox{SNR}_{min}}$ and $T_{min}$ increase significantly for sources that are located far from the antenna  or in the lateral directions.

\section{Conclusion}
\label{conclusion}
In this paper, three important results are proposed, discussed, and assessed through simulation experiments:  $(i)$ Exact conditional and unconditional CRB derivation for near field source localization and its development in non matrix form. The latter reveals interesting features and interpretations not shown by the CRB given in the literature based on an approximate model (i.e., approximate time delay).  $(ii)$ This CRB derivation is generalized to the variable gain case where the exact expression of the received power profile is taken into account. This generalization allows us to investigate the importance of the power profile information in 'adverse' localization contexts (i.e., for lateral lookup directions). \\ $(iii)$ Based on the previous CRB derivation, a new concept of 'localization region' is introduced to better define the space region where the localization quality can meet a target value or otherwise to better tune the system parameters to achieve the target localization quality for a given location region.

\appendices
\label{Annexe}

\section{Proof of Lemma \ref{lm:exact_CRBCvg}}
\label{App1}
A direct calculation of matrix $\tilde{{\mathcal{\textbf{\textit{Q}}}}}$ in (\ref{eq:FIM_bd}) using equation (\ref{eq:FIM_particular}) leads to
\begin{eqnarray*}
	 \tilde{{\mathcal{\textbf{\textit{Q}}}}}= {\left(\begin{array}{c  c}
																						{\bf Q}_{1}   & {\bf Q}_{2} \\																		
																						{\bf Q}_{2}^T & {\bf Q}_{3}   
																						\end{array} \right)}
\end{eqnarray*}
where 
\begin{eqnarray*}
{\bf Q}_{1} &=& {\left( \begin{array}{c  c }
	              f_{\theta \theta} & f_{\theta r} 	\\																		
	              f_{r \theta}      & f_{r r}   
                        \end{array} \right)}\\
{\bf Q}_{2} &=& {\left( \begin{array}{c  c }
	              v_1 & v'_1 	\\																		
              	v_2 & v'_2    
										 \end{array} \right)}\\
{\bf Q}_{3} &=& {\left( \begin{array}{c  c }
                         v_3 diag(\boldsymbol{\alpha}\odot\boldsymbol{\alpha})	 & {\bf 0}_T    	\\																		
                         {\bf 0}_T                         											 & v_3 \I_T   
								         \end{array} \right)}
\end{eqnarray*}
the entries of ${\bf Q}_{1}$ are given by (\ref{ftt}), (\ref{frr}) and (\ref{frt}), $\otimes$ represents the Kronecker product,  $v_1 = \frac{2}{\sigma^2} (\boldsymbol{\gamma} \odot  \boldsymbol{\gamma})^T \boldsymbol {\dot{\tau}}_{\theta}$, $v'_1= \frac{2}{\sigma^2}\boldsymbol{\gamma}^{T}\dot{\boldsymbol{\gamma}}_{\theta}$, \ \ \ \ \ \ \ \ \
$v_2 = \frac{2}{\sigma^2}(\boldsymbol{\gamma} \odot  \boldsymbol{\gamma})^T \boldsymbol {\dot{\tau}}_{r}$, \ \ \ \ \ \ \ $v'_2=\frac{2}{\sigma^2}\boldsymbol{\gamma}^{T}\dot{\boldsymbol{\gamma}}_{r}$,
\\and $v_3 = \frac{2}{\sigma^2}\left\| \boldsymbol{\gamma}  \right\|^2$.

Because the CRB of the range and the angle parameters is equal to the $2 \times 2$ top left sub-matrix of the inverse matrix $\tilde{{\mathcal{\textbf{\textit{Q}}}}}^{-1}$, Schur lemma \cite{schur} can be used and the results will be as 
$
	 \tilde{{\mathcal{\textbf{\textit{Q}}}}}^{-1} = {\left(\begin{array}{c  c }
																								  {\bf Q}^{-1}_{c}    &   {\bf x}  	\\																		
																									{\bf x}             &   {\bf x}  
																									\end{array} \right)}
$ where ${\bf Q}_{c}= {\bf Q}_1- {\bf Q}_{2} .{\bf Q}^{-1}_{3}.{\bf Q}_{2}^T$

After a straightforward computation, one  obtain
\begin{eqnarray*}
{\bf Q}_c= {\left( \begin{array}{c  c }
	f_{\theta \theta} -\frac{\left\| \boldsymbol{\alpha} \right\|^{2}}{v_3}(v^{2}_{1}+v'^{2}_{1})   &  f_{\theta r} -\frac{\left\| \boldsymbol{\alpha} \right\|^{2}}{v_3}(v_1 v_2 + v'_1 v'_2) \\
  f_{r \theta}  -\frac{\left\| \boldsymbol{\alpha} \right\|^{2}}{v_3}(v_1 v_2 + v'_1 v'_2)        &  f_{rr} -\frac{\left\| \boldsymbol{\alpha} \right\|^{2}}{v_3}(v^{2}_{2}+v'^{2}_{2}) 
\end{array} \right)}
\end{eqnarray*}
Now, by comparing this expression of ${\bf Q}_c$ to the expressions in (\ref{eq:Evg_th}), (\ref{eq:Evg_r}) and (\ref{eq:Evg_rth}), one can rewrite
\begin{eqnarray*}
{\bf Q}_c= 2T\tilde{D}_{\mbox{SNR}}{\left( \begin{array}{c  c }
	  E_{vg}(r) &  -E_{vg}(r,\theta)\\
       -E_{vg}(r,\theta) &  E_{vg}(\theta)
\end{array} \right)}
\end{eqnarray*}
leading finally to
\begin{eqnarray*}
\mbox{CRB}^{c}_{vg}(r)        &=&  \frac{E_{vg}(r)}{det({\bf Q}_c)}  \nonumber \\
                              &=& \left(\frac{1}{2TD_{\mbox{SNR}}}\right)\frac{E_{vg}(\theta)}  {E_{vg}(\theta)E_{vg}(r)-E_{vg}(r,\theta)^{2}}      \\
\mbox{CRB}^{c}_{vg}(\theta)   &=&  \frac{E_{vg}(\theta)}{det({\bf Q}_c)} \nonumber \\
                              &=& \left(\frac{1}{2TD_{\mbox{SNR}}}\right)\frac{E_{vg}(r)}       {E_{vg}(\theta)E_{vg}(r)-E_{vg}(r,\theta)^{2}}     \\
\mbox{CRB}^{c}_{vg}(r,\theta) &=&  \frac{E_{vg}(r,\theta)}{det({\bf Q}_c)} \nonumber \\
                              &=& \left(\frac{1}{2TD_{\mbox{SNR}}}\right)\frac{E_{vg}(r,\theta)}{E_{vg}(\theta)E_{vg}(r)-E_{vg}(r,\theta)^{2}}    
\end{eqnarray*}
\section{Proof of Lemma \ref{lm:exact_CRBUvg}}
\label{App2}
For the unconditional case, the considered unknown parameter vector is ${\bf \xi}=(\theta, r, \sigma^{2}_{s}, \sigma^2)^{T}$ which leads to the following $4\times 4$ Fisher Information matrix 
$	\mbox{FIM} = {\left( \begin{array}{c  c }
														 {\bf F}_1       & {\bf F}_2   
														 \\																		
														 {\bf F}^{T}_{2} & {\bf F}_3   
\end{array} \right)}
$ where the $2\times 2$ matrices ${\bf F}_i$ are given by
\begin{eqnarray*}
{\bf F}_1 &=& {\left( \begin{array}{c  c }
	f_{\theta \theta} & f_{\theta r} 
	\\																		
	f_{r \theta} & f_{r r}   
\end{array} \right)}\\
{\bf F}_2 &=& {\left( \begin{array}{c  c }
	f_{\theta \sigma^{2}_{s}} & f_{\theta \sigma^{2}}
	\\																		
	f_{r \sigma^{2}_{s}} & f_{r \sigma^{2}}	 
	\end{array} \right)}\\
	{\bf F}_3 &=& {\left( \begin{array}{c  c }
	f_{\sigma^{2}_{s} \sigma^{2}_{s}} & f_{\sigma^{2}_{s} \sigma^{2}}
	\\																		
	f_{\sigma^{2} \sigma^{2}_{s}} & f_{\sigma^{2} \sigma^{2}}
\end{array} \right)}
\end{eqnarray*}

By using Schur's lemma for matrix inversion \cite{schur}, one can obtain 
$	\mbox{FIM}^{-1} = {\left( \begin{array}{c  c}
																	{\bf L}^{-1} & {\bf G}  
																	\\																		
																	{\bf G}^T    & {\bf H}  
                     \end{array}
             \right)}$ 
where  ${\bf L} = {\bf F}_1 - {\bf F}_2{\bf F}_3^{-1}{\bf F}_2^T = {\left( \begin{array}{c  c}
																																																 u    & x  
																																																 \\																		
																																																 x    & v  
                     																																\end{array}
             																																 \right)}$. ${\bf F}_3$ and ${\bf L}$ are $2\times 2$ matrices and their inverse can be computed easily as
\begin{equation*}
{\bf L}^{-1} = \frac{1}{det}{\left( \begin{array}{c  c }
																									  u    & -x
																									  \\																		
 																								    -x   &  v
																			\end{array} 
														   \right)} = {\left( \begin{array}{c  c }
																									                \mbox{CRB}(r)          &  \mbox{CRB}(\theta,r)
																									               \\																		
 																								                  \mbox{CRB}(r,\theta)   &  \mbox{CRB}(\theta)
																			              \end{array} 
														                 \right)} 		
\end{equation*}															                 											                 
where
\begin{eqnarray}
u     &=& f_{rr} - \frac{1}{det_1}(f_{\theta \sigma^{2}_{s}} c_1 + f_{\theta \sigma^{2}} c_2 ) \label{eq:u}\\
v     &=& f_{\theta \theta} -\frac{1}{det_1}(f_{r \sigma^{2}_{s}} c_1 + f_{r \sigma^{2}} c_2 ) \label{eq:v}\\
x     &=& f_{r \theta}  - \frac{1}{det_1}(f_{r \sigma^{2}_{s}} c_3 + f_{r \sigma^{2}} c_4 )    \label{eq:x}\\
det_1 &=& f_{\sigma^{2} \sigma^{2}} f_{\sigma^{2}_{s} \theta} - f_{\sigma^{2}_{s} \sigma^{2}}f_{\sigma^{2} \theta}          \label{eq:det1}
\end{eqnarray}
\begin{eqnarray}
c_1   &=& f_{\sigma^{2} \sigma^{2}} f_{\sigma^{2}_{s} \theta} - f_{\sigma^{2}_{s} \sigma^{2}}f_{\sigma^{2} \theta}          \label{eq:c1}\\
c_2   &=& f_{\sigma^{2}_{s} \sigma^{2}_{s}} f_{\sigma^{2} \theta} - f_{\sigma^{2} \sigma^{2}_{s}} f_{\sigma^{2}_{s} \theta} \label{eq:c2}\\
c_3   &=& f_{\sigma^{2}_{s} \sigma^{2}_{s}} f_{\sigma^{2}_{s} r} - f_{\sigma^{2}_{s} \sigma^{2}}f_{\sigma^{2} r}            \label{eq:c3}\\
c_4   &=& f_{\sigma^{2}_{s} \sigma^{2}_{s}} f_{\sigma^{2} r} - f_{\sigma^{2} \sigma^{2}_{s}} f_{\sigma^{2}_{s} r}           \label{eq:c4}\\
det   &=& u v - x^2  \label{eq:det}
\end{eqnarray}

Now, it remains only to compute the entries of the FIM by using equation (\ref{eq:FIM_gaussian}) and taking into account that the matrix $\boldsymbol{\Sigma}=\sigma^{2}_{s} \boldsymbol{b}(\theta,r)\boldsymbol{b}(\theta,r)^{H} + \sigma^{2}{\bf I}_{N}$ and its inverse is given as ${\boldsymbol{\Sigma}}^{-1}=\frac{1}{\sigma^{2}}({\bf I}_{N} - \frac{1}{C} \boldsymbol{b}(\theta,r)\boldsymbol{b}(\theta,r)^{H})$ where $C=\frac{1}{\mbox{SNR}} + \left\|\boldsymbol{\gamma} \right\|^2$ and $\boldsymbol{b}(\theta,r)=[\gamma_0, \gamma_1e^{j\tau_1}, \cdots, \gamma_{N-1}e^{j\tau_{N-1}}]^T$.
\\A straightforward (but cumbersome) computation leads to
{\small
\begin{eqnarray*}
f_{\theta \theta}           &=&  \frac{2T}{C^2}(1 - \mbox{SNR} \left\|\boldsymbol{\gamma}\right\|^2) (\boldsymbol{\gamma}^T\dot{\boldsymbol{\gamma}}_{\theta})^2 \\
                            & & - (1 + \mbox{SNR} \left\|\boldsymbol{\gamma}\right\|^2)((\boldsymbol{\gamma} \odot \boldsymbol{\gamma})^T \dot{\boldsymbol \tau}_{\theta})^2 \\ 
                            & & + C \ \mbox{SNR} \left\|\boldsymbol{\gamma}\right\|^2 (\left\|\dot{\boldsymbol{\gamma}}_{\theta}\right\|^2 + (\boldsymbol{\gamma} \odot \boldsymbol{\gamma})^T (\dot{\boldsymbol \tau}_{\theta} \odot \dot{\boldsymbol \tau}_{\theta}))\\
f_{r r}                     &=&   \frac{2T}{C^2}(1 - \mbox{SNR} \left\|\boldsymbol{\gamma}\right\|^2) (\boldsymbol{\gamma}^T \dot{\boldsymbol{\gamma}}_r)^2 \\ 
                            & & - (1+\mbox{SNR} \left\|\boldsymbol{\gamma}\right\|^2)((\boldsymbol{\gamma} \odot \boldsymbol{\gamma})^T \dot{\boldsymbol \tau}_r)^2 \\
                            & & + C \ \mbox{SNR} \left\|\boldsymbol{\gamma}\right\|^2 (\left\|\dot{\boldsymbol{\gamma}}_r \right\|^2+ (\boldsymbol{\gamma} \odot \boldsymbol{\gamma})^T (\dot{\boldsymbol \tau}_r \odot \dot{\boldsymbol \tau}_r))\\
f_{r \theta}                &=& \frac{2T}{C^2}(1 - \mbox{SNR}\left\|\boldsymbol{\gamma}\right\|^2) (\boldsymbol{\gamma}^T \dot{\boldsymbol{\gamma}}_{\theta})(\boldsymbol{\gamma}^T \dot{\boldsymbol{\gamma}}_r) \\
                            & & - (1+\mbox{SNR} \left\|\boldsymbol{\gamma}\right\|^2)((\boldsymbol{\gamma} \odot \boldsymbol{\gamma})^T \dot{\boldsymbol \tau}_{\theta})((\boldsymbol{\gamma} \odot \boldsymbol{\gamma})^T \dot{\boldsymbol \tau}_r) \\ 
                            & & + C \ \mbox{SNR} \left\|\boldsymbol{\gamma}\right\|^2 (\dot{\boldsymbol{\gamma}}_{\theta}^T\dot{\boldsymbol{\gamma}}_r + (\boldsymbol{\gamma} \odot\boldsymbol{\gamma})^T (\dot{\boldsymbol \tau}_{\theta} \odot \dot{\boldsymbol \tau}_r))\\
f_{\sigma^{2}_{s} \sigma^{2}_{s}} &=& \frac{T \left\|\boldsymbol{\gamma}\right\|^4}{\sigma^4 (C \ \mbox{SNR})^2} \\
f_{\sigma^{2} \sigma^{2}}         &=& \frac{T}{\sigma^4 C^2}(N C^2 - \left\|\boldsymbol{\gamma}\right\|^2(2C - \left\|\boldsymbol{\gamma}\right\|^2))\\
f_{\sigma^{2}_{s} \sigma^{2}}     &=& \frac{T \left\|\boldsymbol{\gamma}\right\|^4}{\sigma^4 (C \ \mbox{SNR})^2}\\
f_{\theta \sigma^{2}_{s}}         &=& \frac{2 T \left\|\boldsymbol{\gamma}\right\|^2}{\sigma^2 C^2 \mbox{SNR}} (\boldsymbol{\gamma}^T \dot{\boldsymbol{\gamma}}_{\theta})\\
f_{\theta \sigma^{2}}             &=& \frac{2 T }{\sigma^2 C^2 \mbox{SNR}} (\boldsymbol{\gamma}^T \dot{\boldsymbol{\gamma}}_{\theta})\\
f_{r \sigma^{2}_{s}}              &=& \frac{2 T \left\|\boldsymbol{\gamma}\right\|^2}{\sigma^2 C^2 \mbox{SNR}} (\boldsymbol{\gamma}^T \dot{\boldsymbol{\gamma}}_r)\\
f_{r \sigma^{2}}                  &=& \frac{2 T }{\sigma^2 C^2 \mbox{SNR}} (\boldsymbol{\gamma}^T \dot{\boldsymbol{\gamma}}_r)
\end{eqnarray*}
}
By replacing these entries in equations (\ref{eq:u})-(\ref{eq:det}), we obtain
\begin{eqnarray*}
u     &=& \frac{2 T \mbox{SNR}^2 \left\|\boldsymbol{\gamma}\right\|^2}{(1 + \mbox{SNR} \left\|\boldsymbol{\gamma}\right\|^2)}E_{vg}(r)        \\
v     &=& \frac{2 T \mbox{SNR}^2 \left\|\boldsymbol{\gamma}\right\|^2}{(1 + \mbox{SNR} \left\|\boldsymbol{\gamma}\right\|^2)}E_{vg}(\theta)   \\
x     &=& \frac{2 T \mbox{SNR}^2 \left\|\boldsymbol{\gamma}\right\|^2}{(1 + \mbox{SNR} \left\|\boldsymbol{\gamma}\right\|^2)}E_{vg}(r,\theta) 
\end{eqnarray*}

leading finally to Lemma \ref{lm:exact_CRBUvg} result
\begin{eqnarray*}
\mbox{CRB}^{u}_{vg}(\theta)    &=& \frac{1 + \mbox{SNR} \left\|\boldsymbol{\gamma}\right\|^2}{2 T \mbox{SNR}^2 \left\|\boldsymbol{\gamma}\right\|^2} \frac{E_{vg}(r)}       {E_{vg}(\theta) E_{vg}(r) - E_{vg}(r,\theta)^2} \\
\mbox{CRB}^{u}_{vg}(r)         &=& \frac{1 + \mbox{SNR} \left\|\boldsymbol{\gamma}\right\|^2}{2 T \mbox{SNR}^2 \left\|\boldsymbol{\gamma}\right\|^2} \frac{E_{vg}(\theta)}  {E_{vg}(\theta) E_{vg}(r) - E_{vg}(r,\theta)^2} \\
\mbox{CRB}^{u}_{vg}(r, \theta) &=& \frac{1 + \mbox{SNR} \left\|\boldsymbol{\gamma}\right\|^2}{2 T \mbox{SNR}^2 \left\|\boldsymbol{\gamma}\right\|^2} \frac{E_{vg}(r,\theta)}{E_{vg}(\theta) E_{vg}(r) - E_{vg}(r,\theta)^2} 
\end{eqnarray*}

\ifCLASSOPTIONcaptionsoff
  \newpage
\fi
\bibliographystyle{IEEEbib}
\bibliography{CRBbib}
\end{document}